\documentclass[11pt]{article} 

\usepackage[T1]{fontenc}
\usepackage{color}
\usepackage{graphicx}
\usepackage{hyperref}

\begin{document}


\title{Compression and the origins of Zipf's law of abbreviation}

\author{R. Ferrer-i-Cancho$^{1,*}$, C. Bentz$^2$ \& C. Seguin$^1$ \\
{\small $^1$Complexity \& Quantitative Linguistics Lab, LARCA Research Group,} \\
{\small Departament de Ci\`encies de la Computaci\'o.} \\ 
{\small Universitat Polit\`ecnica de Catalunya,} \\ 
{\small Campus Nord, Edifici Omega Jordi Girona Salgado 1-3.} \\ 
{\small 08034 Barcelona, Catalonia (Spain).} \\
{\small $^2$ Department of Theoretical and Applied Linguistics}, \\
{\small Faculty of Modern \& Medieval Languages (MML).} \\ 
{\small University of Cambridge,} \\ 
{\small Sidgwick Avenue, Cambridge, CB3 9DA, United Kingdom.} \\
~\\
{\small (*) Author for correspondence, rferrericancho@cs.upc.edu}
}


\date{}
	
\maketitle


\begin{abstract}
Languages across the world exhibit Zipf's law of abbreviation, namely more frequent words tend to be shorter. The generalised version of the law - an inverse relationship between the frequency of a unit and its magnitude - holds also for the behaviours of other species and the genetic code. The apparent universality of this pattern in human language and its ubiquity in other domains calls for a theoretical understanding of its origins. To this end, we generalise the information theoretic concept of mean code length as a mean energetic cost function over the probability and the magnitude of the types of the repertoire. We show that the minimisation of that cost function and a negative correlation between probability and the magnitude of types are intimately related. 
\end{abstract}



\section{Introduction}

Zipf's law of abbreviation, the tendency of more frequent words to be shorter \cite{Zipf1949a}, holds in every language for which it was tested \cite{Zipf1949a,Bates2003a,Sigurd2004a,Strauss2007a,Jedlickova2008a,Sanada2008a,Jayaram2009a,
Piantadosi2011,Ferrer2012a} (Fig. \ref{law_of_abbreviation_figure} (a)), suggesting that  language universals are not necessarily a myth \cite{Evans2009a}.   
A generalised version of the law, i.e. a negative correlation between the frequency of a type from a repertoire and its magnitude (e.g., its length, size or duration), has been found in the behaviour of other species \cite{Hailman1985, Ferrer2009g, Semple2010a, Ferrer2012a, Luo2013a} (Fig. \ref{law_of_abbreviation_figure} (b)) and in the genetic code \cite{Naranan2000}. This is strong evidence for a general tendency of more frequent units to be smaller, i.e. less cost-intensive. The robustness and recurrence of this pattern calls for a theoretical understanding of the mechanisms that give rise to it. 

The common interpretation of the law as an indication of the efficiency of language and animal behaviour \cite{Zipf1949a,Ferrer2009b,Semple2010a} suffers from Kirby's problem of linkage, i.e. the lack of a strong connection between potential processing constraints and the proposed universal \cite{Kirby1999a}. 
Here we address the problem of linkage for the law of abbreviation with the help of information theory.
    
Information theory sheds light on the origins of many regularities of natural language, e.g., duality of patterning \cite{Plotkin2000}, Zipf's law for word frequencies \cite{Prokopenko2010a,Ferrer2004e}, 
Clark's principle of contrast \cite{Ferrer2013g}, a vocabulary learning bias in children \cite{Ferrer2013g} and the exponential decay of the distribution of dependency lengths \cite{Ferrer2004b,Alday2015a}. Those examples suggest that the solutions of information theory to communication problems can be informative for natural languages too, though 
efforts in information theory research have been directed towards solving engineering problems, not linguistic problems \cite{Christiansen2015a}. 
 

Here we investigate the law of abbreviation in the light of the problem of compression from standard information theory \cite{Cover2006a,Ferrer2012d}. 
In this context, the mean code length is defined as   
\begin{equation}
L = \sum_{i=1}^V p_i l_i,
\label{mean_length_equation}
\end{equation}
where $p_i$ and $l_i$ are, respectively, the probability and the length in symbols of the $i$-th type of a repertoire of size $V$.
In the case of human language, the types could be words, the symbols could be letters and the repertoire would be a vocabulary.  
Solving the problem of compression provides word lengths that minimise $L$ when the $p_i$'s are given. 
An optimal coding of types by using strings of symbols (under the wide scheme 
of uniquely decipherable codes) satisfies \cite{Cover2006a}
\begin{equation}
l_i = \lceil -\log_N p_i \rceil, 
\label{optimal_coding_equation}
\end{equation}
where $N$ is the size of the alphabet used to code the types.  
Eq. \ref{optimal_coding_equation} is indeed a particular case of Zipf's law of abbreviation.
  
Eq. \ref{mean_length_equation} can also be interpreted as an energetic cost function where the cost of every unit is exactly its length. Based on this assumption, we will generalise the problem of compression in two ways. First, we put forward a cost function $\Lambda$, i.e.
\begin{equation}
\Lambda = \sum_{i=1}^V p_i \lambda_i,
\label{energy_equation}
\end{equation}
where $\lambda_i$ is the energetic cost of the $i$-th type. In his pioneering research, G. K. Zipf
already proposed a particular version of Eq. \ref{energy_equation} to explain the origins of the law of abbreviation using qualitative arguments \cite[p. 59]{Zipf1949a}. Following this line of argument, we address Kirby's problem of linkage \cite{Kirby1999a} showing how the minimisation of $\Lambda$ can produce the law of abbreviation.

We assume that the energetic cost of a unit is a monotonically increasing function of its length $\lambda_i = g(l_i)$. For instance, the energy that is needed to articulate the sounds of a string of length $l_i$ is assumed to increase as $l_i$ increases. 
Second, we generalise $l_i$ as a magnitude (a positive real number).
This way, $l_i$ can indicate not only the length in syllables of a word \cite{Zipf1949a} or the number of strokes of a Japanese kanji \cite{Sanada2008a} but also the duration in time of a vocalisation \cite{Ferrer2012d} or the amount of information of a codon that is actually relevant for coding an amino acid \cite{Naranan2000}. Durations are important in the case of human language because words that have the same length in letters, and even the same number of phonemes, can still have different durations\cite{Gahl2008a}. A review of costs associated with the length or duration of a unit in human language and animal behaviour can be found in \cite{Ferrer2012d}. 

Under these two assumptions $\Lambda$ becomes 
\begin{equation}
\Lambda = \sum_{i=1}^V p_i g(l_i).
\label{energy_equation2}
\end{equation}
$\Lambda$ is equivalent to $L$ when $g$ is the identity function. We assume that $g(l)$ is a strictly monotonically increasing function of $l$. The same assumption has been made for the cost of a syntactic dependency as a function of its length in word order models \cite{Ferrer2013e}.

Assuming that $g$ is the identity function, an equivalence between the minimisation of $\Lambda$ and Zipf's law of abbreviation is suggested by statistical analyses showing that 
any time that $\Lambda$ is significantly small, the correlation between frequency and magnitude is significant (and negative) and {\em vice versa} \cite{Ferrer2012d}. Furthermore, theoretical arguments indicate that 
the law of abbreviation follows from minimum $\Lambda$ when the empirical distribution of type lengths and type frequencies is constant and $g$ is the identity function \cite{Ferrer2012d}. This is indirect evidence of a connection between compression and the law of abbreviation. 

Here we present direct connections between the minimisation of $\Lambda$ and a generalised law of abbreviation:
(1) a link between the minimisation of $\Lambda$ and the maximisation of the concordance with the law of abbreviation using Kendal and Pearson correlation, (2) the relationship between optimal coding and the law of abbreviation through Kendal correlation, and (3) the fact that $\Lambda$ is inherent to the Pearson correlation. In consequence, our research is in the spirit of recent studies on the origins of Zipf's law for word frequencies through optimisation principles \cite{Prokopenko2010a,Ferrer2004e,Dickman2012a}. 

\section{Predicting the law of abbreviation}

\label{law_of_abbreviation_section}

A generalised law of abbreviation is defined often simply as a negative correlation between the frequency of a unit and its magnitude \cite{Ferrer2009g,Semple2010a,Ferrer2012a}. We consider three measures of correlation: Pearson correlation ($r$), Spearman rank correlation ($\rho$) and Kendall rank correlation ($\tau$) \cite{Conover1999a}. While Pearson is a measure of linear association, $\rho$ and $\tau$ are measures of both linear and non-linear association \cite{Gibbons2010a,Embrecths2002a}. 
Alternatives to $r$ are necessary because the functional dependency between frequency and length is modelled by means of non-linear functions \cite{Strauss2007a} and the actual $g$ may not be linear. 
  
$r$ and $\rho$ have been used in previous research on the generalised law of abbreviation \cite{Ferrer2009g,Semple2010a}. Here we additionally introduce $\tau$ \cite{Conover1999a}. 
The non-parametric approach offered by $\rho$ and $\tau$ allows one to remain agnostic about the actual functional dependency between frequency and magnitude \cite{Strauss2007a} and avoids common problems of assuming concrete functions for linguistic laws \cite{Altmann2015a, Ferrer2012h}.
  
\subsection{The relationship between $\Lambda$ and $\tau$}

\label{relationship_between_mean_energetic_cost_and_Kendall_tau_section}

Here we will unravel 
a strong dependency between the minimisation of $\Lambda$ and the minimisation of Kendall's $\tau$, a rank measure of correlation between $p_i$ and $l_i$, under two different conditions: multiset constancy and minimum $\Lambda$. 
We have multiset constancy when the multiset of $p_i$'s and the multiset of 
$\lambda_i$'s are constant. This condition has been used to test the significance of 
$\Lambda$ \cite{Ferrer2012d}, and is the central assumption of correlation tests such as the ones used to test for the law of abbreviation \cite{Ferrer2012h}.

$\tau$ is based on the concept of concordant and discordant pairs. In our case, 
$(p_i, l_i)$ and $(p_j,l_j)$ are
\begin{itemize}
\item
Concordant if either $p_i > p_j$ and $l_i > l_j$ or $p_i < p_j$ and $l_i < l_j$
\item
Discordant if either $p_i > p_j$ and $l_i < l_j$ or $p_i < p_j$ and $l_i > l_j$
\item
Non concordant if $p_i = p_j$ or $l_i = l_j$.
\end{itemize}
$n_c$ and $n_d$ are defined, respectively, as the number of concordant and discordant pairs. $n_0$ is the total number of pairs, i.e. 
\begin{equation}
n_0 = \frac{V(V-1)}{2}.
\end{equation}
Then Kendall's $\tau$ is defined as a normalised difference between $n_c$ and $n_d$, i.e.
\begin{equation}
\tau = \frac{n_c - n_d}{n_0}.
\label{tau_equation}
\end{equation}
If the agreement with Zipf's law of abbreviation was perfect, i.e. if we had only discordant pairs, then we would have $\tau = -1$.

We will investigate the consequences of choosing a pair of indices at random, $i$ and $j$ ($i\neq j$) and swapping either $p_i$ and $p_j$ or $l_i$ and $l_j$. A prime will be used to indicate the value of a quantity or measure after the swap. For instance, $\tau'$ and $l_i'$ indicate, respectively, the value of $\tau$ and that of $l_i$ after the swap. $\Delta_\tau = \tau'-\tau$ and $\Delta_\Lambda = \Lambda'-\Lambda$ indicate the discrete derivative of $\tau$ and $\Lambda$, respectively. For instance, $\Delta_\tau <0$ indicates that the concordance with the law of abbreviation increases after one swap.
If $p_i$'s are real probabilities (not frequencies from a sample) and the $l_i$'s are durations \cite{Gahl2008a,Semple2010a} (not discrete lengths), ties are unlikely. For this reason we assume that there are no ties (and therefore non concordant pairs are missing) to investigate the relationship between $\Lambda$ and $\tau$. This has the further advantage of simplifying the mathematical arguments.   
A careful analysis shows that (see supplementary online information for further details)
\begin{itemize}
\item
if the pair $(p_i,l_i)$ and $(p_j, l_j)$ is concordant, then $\Delta_\Lambda, \Delta_\tau < 0$. 
\item
if the pair $(p_i,l_i)$ and $(p_j, l_j)$ is discordant, then $\Delta_\Lambda, \Delta_\tau > 0$.
\end{itemize}
The results above can be summarised as    
\begin{equation}
\Delta_\Lambda \Delta_\tau > 0.
\end{equation}
This result means that if one of the two changes (e.g., $\Lambda$), the other (e.g., $\tau$) also changes in the same direction (and {\em vice versa}).  


It is easy to show 
that minimum $\Lambda$ implies $\tau \leq 0$ (even if there are ties of probability or magnitude). The fact that removing a concordant pair (by swapping) always decreases $\Lambda$ allows us to conclude that $n_c = 0$ is a necessary condition for minimum $\Lambda$. Applying $n_c = 0$ to Eq. \ref{tau_equation} and knowing that $n_c$ and $n_0$ are positive, one concludes that $\tau \leq 0$ with equality if and only if $n_d = 0$. The latter condition can be refined taking into count the optimal coding with non-singular codes with an alphabet of size $N$ (supplementary online information) and assuming that there are no probability ties. In that case, $n_d = 0$ is equivalent to $V \leq N$. These arguments strengthen and generalise previous results on the need of the law of abbreviation in case of optimal coding assuming that (1) $g$ is the identity function and that (2) swaps are applied to types that are consecutive in an ordering of types by decreasing probability \cite{Ferrer2012d}.

Let us come back to the general framework of standard information theory under the scheme of non-singular codes, namely two different types cannot be assigned the same string. In that case, optimal coding (minimum $\Lambda$ when the $p_i$'s are given) can be achieved with the following procedure: 
\begin{enumerate}
\item
Generate the sequence of the $V$ shortest strings that can be produced with an alphabet of size $N$. This gives a sequence of strings from length $l = 1$ to length $l_{max}$ (for length $
l = l_{max}$ it might be necessary to choose some of the $N^{l_{max}}$ strings of length $l_{max}$ arbitrarily).
\item
Sort the sequence by increasing length (and within each length following lexicographic order). In case of a binary alphabet ($N = 2$), and $V = 10$, this yields the sequence
0, 1, 00, 01, 10, 11, 000, 001, 010, 011. 
\item
Assign the $i$-th string of the sequence above to the $i$-th most probable type (in case of a tie in $p_i$'s, sort the types involved arbitrarily). 
\end{enumerate}
It is easy to see that the coding procedure minimises $\Lambda$. First, notice that it satisfies the requirement that $n_c = 0$. Second, notice that it involves the shortest strings possible and is therefore optimal (a detailed proof is available in the supplementary online information). 

If $p_i$ is the probability of the $i$-th most probable type, the optimal procedure above yields (see supplementary online information)
\begin{equation}
l_i = \left\{
         \begin{array}{ll}     
         \left\lceil \log_N \left(\frac{N-1}{N}i + 1 \right) \right\rceil & \mbox{~for~} N > 1 \\ 
         i & \mbox{~for~} N = 1. \\
         \end{array}
      \right.
\label{optimal_non_singular_coding_equation}
\end{equation}
When $N$ and $i$ are sufficiently large,
\begin{equation}
l_i \approx \log_N i.
\end{equation} 
Notice that Eq. \ref{optimal_non_singular_coding_equation} relates the rank of a type (according to its probability) with its length for optimal non-singular codes. Interestingly, that equation can be regarded as an analogue of Eq. \ref{optimal_coding_equation}, which relates length and probability for optimal uniquely decipherable codes. Both Eq. \ref{optimal_coding_equation} and Eq. \ref{optimal_non_singular_coding_equation} mean that length tends to grow as probability rank increases and thus lead to a negative correlation between probability and magnitude. Therefore, a negative correlation between probability and magnitude is expected both under multiset constancy and also when there is freedom to assign any magnitude. 

Bear in mind that removing all concordant pairs can lead to a drastic reduction of $\Lambda$ but does not warrant that the coding is optimal in an information theoretic sense.
Suppose that magnitudes are string lengths (in symbols), and that there are neither probability ties nor length ties, i.e. $n_c - n_d = n_0$. After removing all concordant pairs by swapping we get $n_c = 0$, and thus Eq. \ref{tau_equation} gives $\tau = -1$, the strongest negative correlation possible. However, optimal coding with discrete units implies length ties for $V > 2$ (supplementary online information).

\subsection{The relationship between $\Lambda$ and $\rho$}

\begin{figure}[!htb] 
\begin{center}
\includegraphics[scale = 0.45]{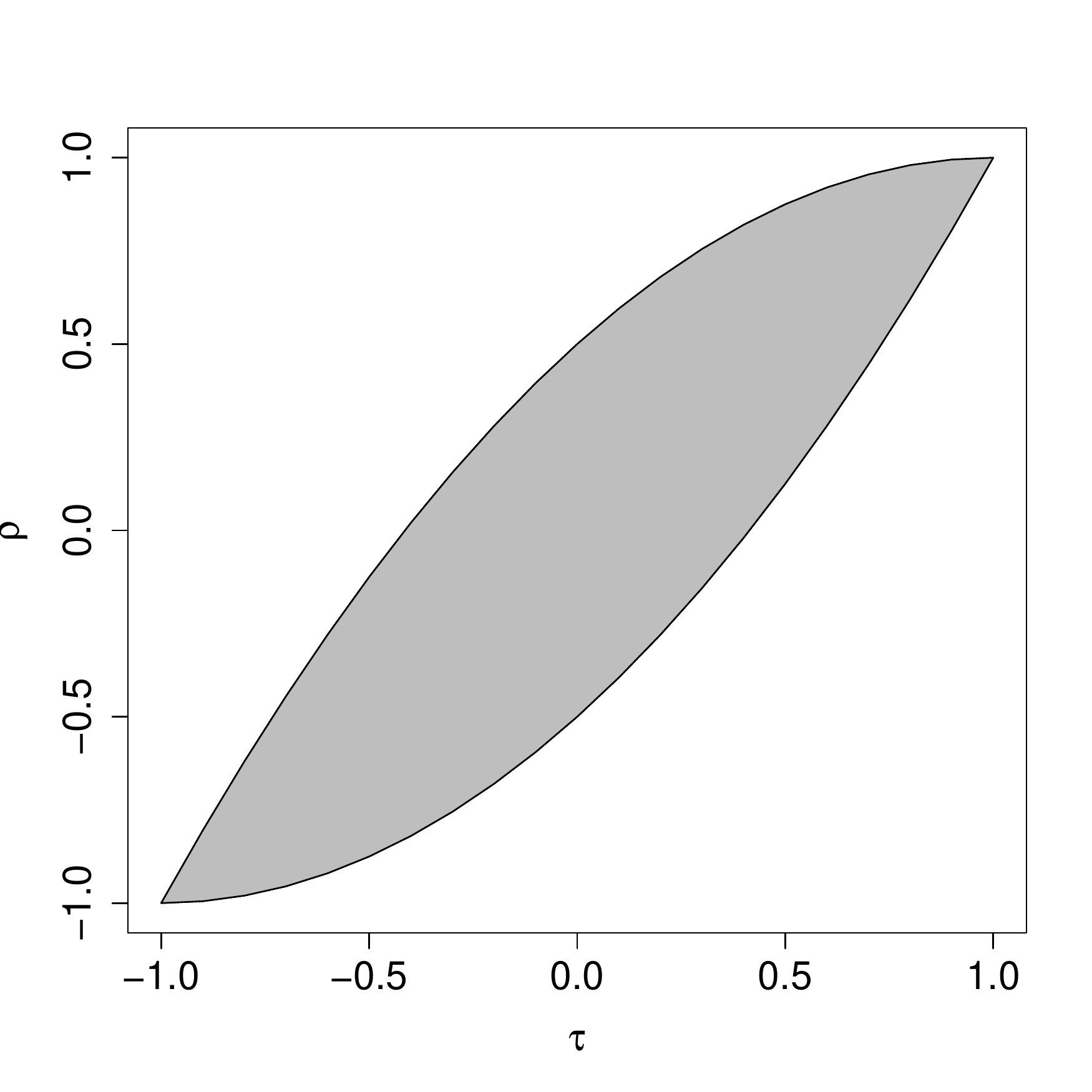}
\end{center}
\caption{\label{bounding_Spearman_correlation_figure} 
The gray region covers the values of Spearman $\rho$ satisfying Eq. \ref{bounding_Spearman_correlation_equation} as a function of Kendall $\tau$.    
}
\end{figure}

Indirect relationships between $\rho$ and $\Lambda$ follow from those between $\rho$ and $\tau$. On the one hand, $\rho$ satisfies \cite{Daniels1950a} 
\begin{equation}
\frac{1}{2}(3\tau - 1) \leq \rho \leq \frac{1}{2}(1 + 3\tau).
\end{equation}
On the other hand, $\rho$ satisfies \cite{Durbin1951a}
\begin{equation}
\frac{(1 + \tau)^2}{2} - 1 \leq \rho \leq 1 - \frac{(1 - \tau)^2}{2},
\label{bounding_Spearman_correlation_equation}
\end{equation}
as illustrated in Fig. \ref{bounding_Spearman_correlation_figure}.
See \cite{Fredricks2007a} for further relationships between $\rho$ and $\tau$. 

\subsection{The relationship between $\Lambda$ and Pearson's $r$}

\label{mean_energetic_cost_and_Pearson_correlation_section}  

The Pearson correlation between the probability of a unit ($p$) and its energetic cost ($\lambda$) is 
\begin{equation}
r = \frac{E[p\lambda] - E[p]E[\lambda]}{\sigma[p] \sigma[\lambda]}, \label{Pearson_correlation_equation}
\end{equation} 
where $E[x]$ and $\sigma[x]$ are, respectively, the expectation and the standard deviation of a random variable $x$. $E[x]$ can be regarded as the average value of $x$ obtained when drawing types uniformly at random from the repertoire. 
Knowing 
\begin{eqnarray}
E[p\lambda] = \frac{1}{V} \sum_{i=1}^V p_i \lambda_i = \frac{\Lambda}{V} \\
E[p]  = \frac{1}{V} \sum_{i=1}^V p_i = \frac{1}{V},
\end{eqnarray}
Eq. \ref{Pearson_correlation_equation}, can be expressed as
\begin{equation}
r = \frac{\Lambda - E[\lambda]}{V\sigma[p] \sigma[\lambda]}. \label{Pearson_correlation2_equation}
\end{equation}  
Therefore, $r$ is a function of $\Lambda$.    

Note that in quantitative linguistics, the term {\em type} is used to refer to a string of symbols and the term {\em token} is used to refer to an occurrence of a type \cite{Covington2010a}. The term type is used with the same meaning in quantitative studies of animal behaviour \cite{McCowan1999} or in Mandelbrot's pioneering work \cite{Mandelbrot1966}. 
$E[\lambda]$ and $\sigma[\lambda]$ are, respectively, the mean cost and the standard deviation of the cost of the types.

Now, let us assume a constancy condition: $V$, $E[\lambda]$, $\sigma[p]$ and $\sigma[\lambda]$ are constant (see the supplementary online information for a justification). If that simplifying condition holds, Eq. \ref{Pearson_correlation2_equation} indicates that the minimisation of 
$\Lambda$ is equivalent to the minimisation of $r$, which in turn maximises the concordance with the law of abbreviation because $g(l)$ is a monotonically increasing function of $l$. 
The same conclusion can be reached with a simpler but less general constancy condition, namely, multiset constancy. 

$r<0$ can be regarded as concordance with the law of abbreviation. 
To know if $r<0$, it is not necessary to actually calculate $r$ with Eq. 
\ref{Pearson_correlation2_equation}. On the right hand side of this equation, the denominator is positive (since $V$ and the standard deviations are positive). Hence, the sign of $r$ is determined by the sign of the numerator. Therefore, 
$r < 0$ if and only if 
\begin{equation}
\Lambda = \sum_{i = 1}^V p_i \lambda_i < E[\lambda] = \frac{1}{V}\sum_{i = 1}^V \lambda_i, 
\label{sign_of_Pearson_correlation_equation}
\end{equation}
i.e. the expected energetic cost of types when selecting them according to their probability ($p$) is smaller than the expected energetic cost of types picking them uniformly at random from the repertoire.
Eq. \ref{sign_of_Pearson_correlation_equation} tell us that a negative sign of $r$ is equivalent to a mean energetic cost of {\em tokens} that does not exceed the mean energetic cost of {\em types}. 

A limitation of the connection between $r$ and $\Lambda$ above is not only the validity of the constancy condition but also that $r$ is a measure of linear association. {A priori}, we do not know if the relationship between $p$ and $\lambda$ is linear. For this reason it is vital to explore a connection between measures of correlation that can capture non-linear dependencies.  

\section{Discussion}

Assuming some constancy conditions on probabilities and magnitudes, we have shown an intimate relationship between the minimisation of $\Lambda$ and the minimisation of various measures of correlation between the probability of a type and its magnitude. Notice that minimisation of the correlation is equivalent to the maximisation of the concordance with the law of abbreviation. This potentially explains the ubiquity of a generalised version of Zipf's law of abbreviation in human language and also in the behaviour of other species. 

More specifically, we have shown that Pearson's $r$ contains $\Lambda$ in its definition (Eq. \ref{Pearson_correlation_equation}). A straightforward relationship between a function to optimise and correlation is also found in methods for community detection in networks (see supplementary online information for further details).
Our mathematical results 
further shed light on previous results on the law of abbreviation involving $r$. 

First, $r$ has been used to investigate a generalised version of Zipf's law of abbreviation in dolphins surface behavioural patterns \cite{Ferrer2009g} and the vocalisations of Formosan macaques \cite{Semple2010a}. The magnitude of dolphin surface behavioural patterns was measured in elementary behavioural units while the magnitude of Formosan macaque vocalisations was measured by their duration. 
The mathematical connections presented in Section \ref{mean_energetic_cost_and_Pearson_correlation_section} predict 
a significantly low mean cost $\Lambda$ for Formosan macaques \cite{Semple2010a} and dolphins via the significant negative $r$ found in both species \cite{Semple2010a,Ferrer2009g}. Only two assumptions are required: multiset constancy for both the correlation test and the test of significance of $\Lambda$, and that the energetic cost of a signal is proportional to its magnitude.
This prediction is confirmed by the analysis of the significance of $\Lambda$ in dolphin surface behavioural patterns and the vocalisation of Formosan macaques \cite{Ferrer2012d}. 

Second, the same arguments predict a significant negative Pearson correlation between frequency and magnitude from the significantly low $\Lambda$ that has been found in various languages \cite{Ferrer2012d}, although that correlation was not investigated for them.

In spite of the predictive power of the minimisation of $\Lambda$, we do not mean that the law of abbreviation is inevitable. Exceptions are known in other species \cite{Ferrer2012a, Luo2013a}. This is not surprising from the perspective of information theory. Solving the problem of compression is in conflict with the problem of transmission: redundancy must be added in a controlled fashion to combat noise in the channel \cite[p. 184]{Cover2006a}. Consistently, the law of abbreviation prevails in short range communication \cite{Ferrer2012a, Luo2013a}. 

\subsection{Compression versus random typing}

Simple mechanisms such as random typing \cite{Conrad2004a,Miller1957,Miller1963,Li1992b}, can reproduce the law of abbreviation \cite{Ficken1978a}. Random typing models produce "words" by concatenating units, one of them behaving as "word" delimiter.
Some researchers regard random typing as not involving any optimisation at all \cite{Miller1963,Li1992b}. However, its conceivable that the dynamical rules of random typing arise or are reinforced or stabilised by compression, given 
\begin{itemize}
\item
The equivalence between the law of abbreviation and compression outlined above.
\item
The optimality of the nonsingular coding scheme in the definition of random typing models (see supplementary online information for further details).
\end{itemize}
In that case, random typing could be seen as a special manifestation of compression. Another connection between optimisation and random typing is that stringing subunits to form "words" as in random typing is a convenient strategy to combat noise in communication \cite{Plotkin2000}.

Having said this, it is unlikely that random typing is the mediator between compression and the law of abbreviation in human languages.
A serious limitation of random typing models is that the probability of a word is totally determined by its composition. In simple versions of the model \cite{Miller1957,Miller1963,Li1992b}, the probability of a word is determined by its length (the characters constituting the words are irrelevant), i.e. the length of a word ($l$) is a decreasing linear function of its probability ($p$)
\begin{equation}
l = a \log p + b, \label{law_of_abbreviation_in_random_typing_equation}
\end{equation}
where $a$ and $b$ are constants ($a$ < 0). To see it, notice that 
the probability of a "word" $w$ is \cite[p. 838]{Ferrer2009a}
\begin{equation}
p(w) = \left( \frac{1-p_s}{N} \right) ^{l} \frac{p_s}{(1-p_s)^{l_0}},
\label{probability_of_a_word_in_random_typing_equation}
\end{equation}
where $l$ is the length of $w$, $p_s$ is the probability of producing the word delimiter, $N$ is the size of the alphabet that the words consist of ($N$ > 1) and $l_0$ is the minimum word length ($l_0 \geq 0$). 
Eq. \ref{probability_of_a_word_in_random_typing_equation} allows one to express $l$ as a function of $p(w)$. Rearranging the terms of Eq. \ref{probability_of_a_word_in_random_typing_equation}, taking logarithms, and replacing $p(w)$ by $p$,  one recovers Eq. \ref{law_of_abbreviation_in_random_typing_equation} with 
\begin{equation}
a = \left( \log \frac{1-p_s}{N} \right)^{-1}
\end{equation} 
and 
\begin{equation}
b = a \log \frac{(1-p_s)^{l_0}}{p_s}.
\end{equation}

Another limitation of random typing is that it is not a plausible model for human language from a psychological perspective \cite{Ferrer2009b} and also from a social perspective: the "words" produced by random typing are not constrained by a predetermined vocabulary of words whose meanings have been agreed upon by social interaction among individuals as in human language \cite{Baronchelli2005a}. 

Also, comparing the statistical properties of random typing against real languages reveals striking differences: 
\begin{itemize}
\item
The distribution of "word" frequencies deviates from that of actual word frequencies significantly \cite{Ferrer2009b}. 
\item 
While random typing yields a geometric distribution of "word" lengths, the actual distribution of word lengths is not a monotonically decreasing function of $l$ \cite{Newman2004a,Manin2009a}.
\item 
In the simple versions of the model, words of the same length are equally likely, a property that real languages do not satisfy \cite{Leopold1998,Ferrer2001c}.
\end{itemize}
Another challenge for random typing are homophones: words that have the same composition (i.e. the same sequence of phonemes, and thus, the same length in phonemes) but can have different frequencies. Interestingly, given a pair of homophones the more frequent one tends to have a shorter duration in time \cite{Gahl2008a}. This is impossible in random typing but explainable by the minimisation of $\Lambda$. 
However, random typing might be relevant for the finding of the law of abbreviation in bird song \cite{Ficken1978a}, where the need for social consensus about the meaning of a song is missing or secondary while pressure for song diversity is crucial to maximise the chances of mating \cite{Catchpole1995a}.

\subsection{From Zipf to standard information theory}

At a more general level, our contributions can be interpreted in two directions. First, we have made a step forward in formalising Zipf's view (e.g., Zipf's minimum equation) \cite{Zipf1949a} with information-theoretic rigour. Second, we have contributed to expand standard information theory beyond uniquely decipherable codes. We have presented an optimal coding procedure under the broader class of non-singular codes, and shown that the minimisation of cost leads to a negative correlation between the probability of a type and its magnitude (a generalised law of abbreviation) under wide conditions. These results are crucial for research into natural communication systems. While standard information theory is focused on uniquely decipherable codes \cite{Cover2006a,Berstel2009a,Borda2011a,Gray2011a}, real languages do not fit that scheme: given a string of letters, there might be more than one way of breaking it into words. We hope that our work on optimal coding helps to change the view that information theory does not contribute to understanding natural language problems \cite{Christiansen2015a}. Further linguistic applications of our theoretical framework beyond the law of abbreviation are presented in the supplementary online information.

\subsection{Causality}
 
A challenge for our theoretical arguments is the extent to which the minimisation of $\Lambda$ is a causal force for the emergence of Zipf's law of abbreviation. The apparent universality of the law of abbreviation in languages and the multiple theoretical connections between compression and the law suggest that the minimisation of $\Lambda$ is indeed a causal force. Additional support for compression as cause may come from the investigation of its predictions in
grammaticalisation, a process of language change by which words become progressively more frequent and shorter \cite{Heine2007, Bybee1994}. In the worst case (i.e. compression is not the driving force), compression would still illuminate the optimality of the law of abbreviation and its stability once it was reached through a mechanism unrelated to compression.   

\section*{Authors' contributions}

RFC conceived the mathematical work. RFC and CB drafted the manuscript.
RFC and CS performed the mathematical work. CS and CB revised critically the article. All authors gave final approval for publication.

\section*{Acknowledgements}

\begin{figure}[!htb] 
\begin{center}
\includegraphics[scale = 0.6]{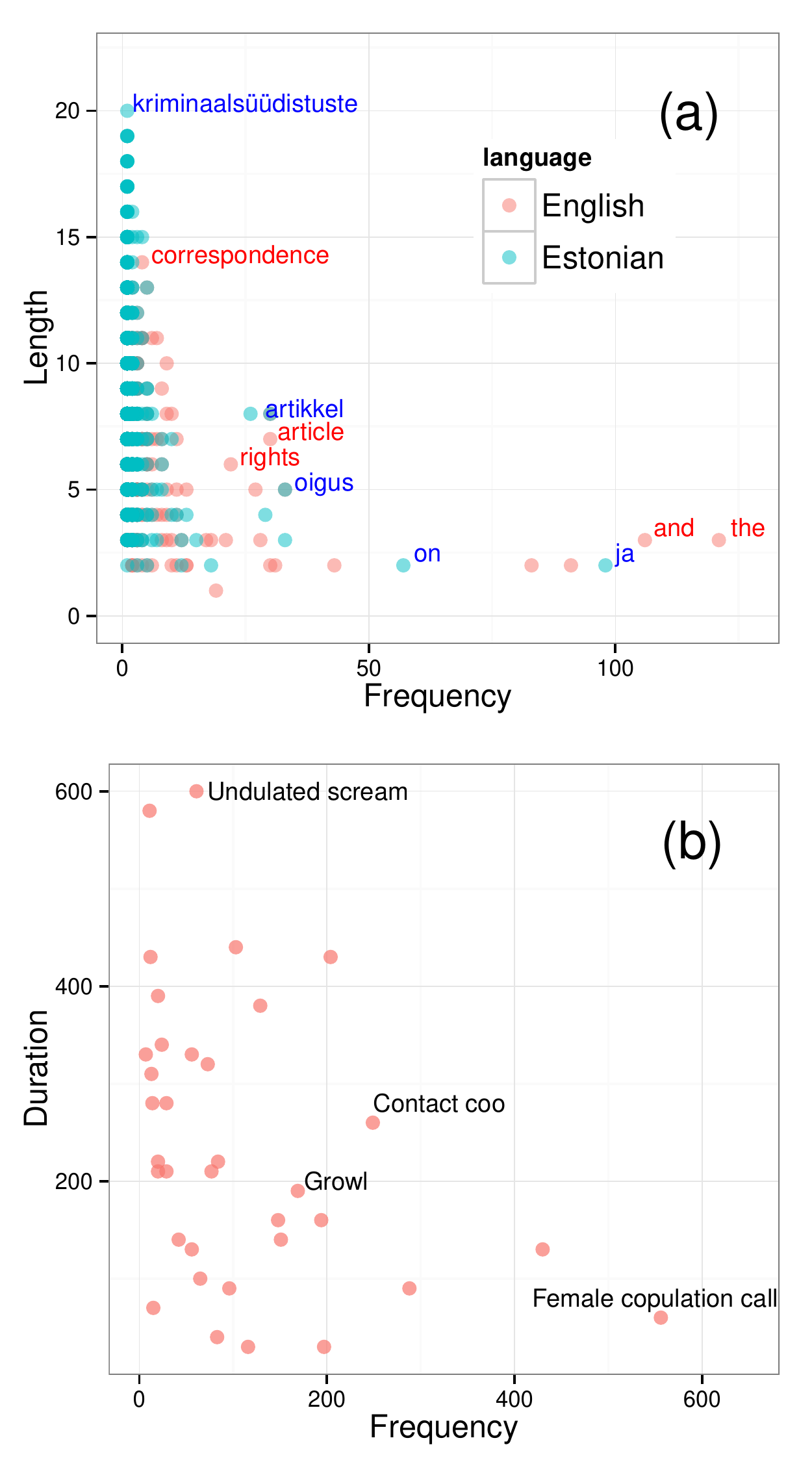}
\end{center}
\caption{\label{law_of_abbreviation_figure} 
The law of abbreviation in human languages and Formosan macaques. (a) The relationship between word frequency and word length (in characters) for the English and Estonian translations of Universal Declaration of Human Rights (\protect \url{http://www.unicode.org/udhr/}). The Estonian words chosen are 
{\em ja} (and), {\em on} (the), {\em oigus} (right), {\em artikkel} (article) and {\em kriminaals\"u\"udistuste} (criminal prosecutions). 
(b) The relationship between call type frequency call type mean duration (in ms) in Formosan macaques (data borrowed from \cite{Semple2010a}). }
\end{figure}

	We thank M. E. J. Newman for indicating the connection between modularity and correlation, and S. Semple, G. Agoramoorthy and M.J. Hsu for the opportunity to use the macaque data for Fig. \ref{law_of_abbreviation_figure} (b). Special thanks to M. Arias and L. Debowski for helping us to strengthen some of the mathematical proofs. We are also grateful to M. Arias, A. Arratia, N. Ay, L. Debowski, M. Gustison, A. Hern\'andez-Fern\'andez and S. Semple for valuable discussions.   
RFC is funded by the grants 2014SGR 890 (MACDA) from AGAUR (Generalitat de Catalunya) and also
the APCOM project (TIN2014-57226-P) from MINECO (Ministerio de Economia y Competitividad).
CB was funded by an Arts and Humanities Research Council (UK) doctoral grant and Cambridge Assessment (reference number: RG 69405), as well as a grant from the Cambridge Home and European Scholarship Scheme. At a later stage, CB was also supported by the EVOLAEMP project and the DFG Center for Advanced Studies \textit{Words, Bones, Genes, Tools} at the University of T\"{u}bingen.
CS is funded by an Erasmus Mundus master scholarship granted by the Education, Audiovisual and Culture Executive Agency of the European Commission.

\bibliographystyle{vancouver}

\bibliography{../law_of_abbreviation/biblio,Ramon}

\pagebreak

\pagebreak

\pagebreak

\noindent {\bf {\Large SUPPLEMENTARY ONLINE INFORMATION}}

\appendix

\section{Optimal coding}

We have put forward a cost function $\Lambda$, i.e.
\begin{equation}
\Lambda = \sum_{i=1}^V p_i \lambda_i,
\label{energy_supplementary_equation}
\end{equation}
where $p_i$ and $\lambda_i$ are, respectively, the probability and the energetic cost of the $i$-th type. We have assumed that the energetic cost of a type is a monotonically increasing function of its magnitude $l_i$, i.e. $\lambda_i = g(l_i)$. When $g(l_i) = l_i$ and $l_i$ is the length in symbols of the alphabet, $\Lambda$ becomes $L$, the mean code length of standard information theory \cite{Cover2006a}. 

Here investigate the minimisation of $\Lambda$ when the $p_i$'s are given. 

\subsection{Optimal coding without any constraint}

The solution to the minimisation of $\Lambda$ when no further constraint is imposed is that all types have minimum magnitude, i.e.
\begin{equation}
l_i = l_{min} \mbox{~for~} i = 1, 2,..., V.
\label{trivial_minimization_equation}
\end{equation}
If $l_i$ is the length of the $i$-th symbol with $l_i \geq 1$, then $l_{min} = 1$.  
If confusion between types has to be avoided, the unconstrained minimisation of $\Lambda$ implies that $N$, the size of the alphabet used to build strings of symbols, cannot be smaller than $V$. 


\subsection{Optimal coding with nonsingular codes}

Standard information theory bears on the elementary assumption that different types cannot be represented by the same string of symbols \cite{Cover2006a}. 
Under the wide scheme 
of uniquely decipherable codes, standard information tell us that the minimisation of $L$ leads to \cite{Cover2006a}
\begin{equation}
l_i \propto \lceil -\log p_i \rceil, 
\label{optimal_coding_supplementary_equation}
\end{equation} 
which is indeed a particular case of Zipf's law of abbreviation. 

Here we investigate the optimal coding using nonsingular codes, a superset of uniquely decipherable codes \cite[p. 106]{Cover2006a}. The function to minimise is $\Lambda$, a generalisation of $L$ the mean code length of standard coding theory \cite{Cover2006a}. 
 
We consider the set of all the strings of symbols that can be built with an alphabet of size
$N$. Suppose that we sort the strings by increasing length  (the relative ordering of strings of the same length is arbitrary), thus the strings in positions $1$ to $N$ have length $1$, the strings in positions $N+1$ to $N+N^2$ have length 2, and so on. 
Suppose that types to be coded are sorted by decreasing probability, i.e. 
\begin{equation}
p_1 \geq p_2 \geq ... \geq p_V, \label{ordering_of_probabilities_equation} 
\end{equation}
being $p_i$ the probability of the $i$-th type. Suppose that we assign the $i$-th string to the $i$-th type for $i = 1, 2, ... V$ . An example of this coding are random typing models \cite{Miller1963,Mandelbrot1966,Li1992b,Conrad2004a}.  
We will show that this coding method, which we refer to as method A is optimal.  
We will proceed in two steps. First, recall that minimum $\Lambda$ requires $n_c = 0$, where $n_c$ is the number of concordant pairs (as explained in the main article). Recall also that the pair $(p_i,l_i)$ and $(p_j, l_j)$ is concordant if and only if $p_i < p_j$ and $l_i < l_j$ or $p_i > p_j$ and $l_i > l_j$. Notice that, by definition, method A produces no concordant pairs.
Second, suppose that 
\begin{equation}
\Lambda^x = \sum_{i=1}^V p_i \lambda_i^x, 
\end{equation}
where $\lambda_i^x$ is the cost of the $i$-th type according to some coding method $x$. 
We will show by induction on $V$ that for any alternative method B that is based on nonsingular codes with $n_c = 0$, $\Lambda^B \geq \Lambda^A$. 
\begin{itemize}
\item
{\em Setup}. 
Since method A only produces pairs that are either concordant or discordant and probabilities obey Eq. \ref{ordering_of_probabilities_equation}, we have
\begin{eqnarray}
\lambda_1^A \leq ... \leq \lambda_i^A ... \leq \lambda_V^A.
\end{eqnarray}
Suppose that there are no probability ties. Then 
\begin{equation}
\lambda_1^B \leq ... \leq \lambda_i^B ... \leq \lambda_V^B
\label{increasing_magnitudes_equation}
\end{equation}
follows immediately.
In case of probability ties, Eq. \ref{increasing_magnitudes_equation} may not hold. Suppose 
$V = 3$, $p_1 = p_2 = p_3$ and $l^B_1=2$, $l^B_2=l^B_3=1$. This coding lacks concordant pairs but does not satisfy Eq. \ref{increasing_magnitudes_equation}. However, any coding produced by method B can be converted into one that satisfies Eq. \ref{increasing_magnitudes_equation} with the same $\Lambda$ by sorting all magnitudes increasingly in every probability tie. We assume that the codings produced by method B have been rearranged in this fashion. This is crucial for the inductive step. 
\item
{\em Basis}: $V = 1$. Then 
\begin{eqnarray}
  \Lambda^A = p_1 \lambda_1^A = p_1 g(1) \label{base_case_B_equation} \\
  \Lambda^B = p_1 \lambda_1^B = p_1 g(l_1). \label{base_case_A_equation}
\end{eqnarray}
Recalling that $g$ is a strictly monotonically increasing function, Eqs. \ref{base_case_B_equation} and \ref{base_case_A_equation} indicate that the condition $\Lambda^B \geq \Lambda^A$ is equivalent to $l_1 \geq 1$, which is trivially true by the definition of $l_i$.
\item
{\em Inductive hypothesis}. If Eq. \ref{increasing_magnitudes_equation} holds then
\begin{equation} 
\sum_{i=1}^{V} p_i \lambda_i^B \geq \sum_{i=1}^{V} p_i \lambda_i^A
\end{equation}
\item
{\em Inductive step}
We want to show that
\begin{equation} 
\sum_{i=1}^{V+1} p_i \lambda_i^B \geq \sum_{i=1}^{V+1} p_i \lambda_i^A
\label{inductive_step_equation}
\end{equation}
when  
\begin{equation}
\lambda_1^B \leq ... \leq \lambda_i^B ... \leq \lambda_{V+1}^B.
\label{extended_increasing_magnitudes_equation}
\end{equation}
Eq. \ref{inductive_step_equation} is equivalent to
\begin{equation}
S + p_{V+1} (\lambda_{V+1}^B - \lambda_{V+1}^A) \geq 0, \label{inequality_to_prove_equation}
\end{equation}
with  
\begin{equation} 
S = \sum_{i=1}^{V} p_i (\lambda_i^B - \lambda_i^A).
\end{equation}
Let us define $l_i^x$ as the magnitude of the $i$-th type according to some coding method $x$.
To show that Eq. \ref{inequality_to_prove_equation} holds, it suffices to show that 
\begin{equation}
l_{V+1}^B \geq l_{V+1}^A, \label{simple_inequality_to_prove_equation}
\end{equation} 
because $S \geq 0$ by the induction hypothesis (notice that if Eq. \ref{extended_increasing_magnitudes_equation} holds then Eq. \ref{increasing_magnitudes_equation} also holds), $p_{V+1}$ is positive by definition and 
$\lambda_i^x = g(l_i^x)$, where $g$ is a strictly monotonically increasing function. 
Notice that if 
\begin{equation}
l_{V+1}^B < l_{V+1}^A
\label{impossible_condition_equation}
\end{equation}
then method B would not employ nonsingular codes. To see it, notice that $l_{V+1}^A$ is the smallest integer such that 
\begin{equation}
V \leq \sum_{l=1}^{l_{V+1}^A} N^l, 
\end{equation}
where $N^l$ are all the strings of length $l$ that can be produced. Thus, method B must assign the same string to different types when Eq. \ref{impossible_condition_equation} holds.
\end{itemize}  

We aim to derive the relationship between the rank of a type (defined according to is probability) and its length in case of optimal non-singular codes for $N \geq 1$. Suppose that $p_i$ is the probability of the $i$-th most probable type and that $l_i$ is its length. The largest rank of types of length $l$ is  
\begin{equation}
i = \sum_{k=1}^l N^k.
\end{equation}
When $N > 1$, we get
\begin{equation}
i = \frac{N(N^l-1)}{N-1}
\end{equation}
and equivalently 
\begin{equation}
N^l = \frac{N-1}{N}i + 1.
\end{equation}
Taking logs on both sides of the equality, one obtains
\begin{equation} 
l = \frac{\log \left(\frac{N-1}{N}i + 1 \right)}{\log(N)}.
\end{equation}
The result can be generalised to any rank of types of length $l$ as
\begin{equation} 
l = \left\lceil \frac{\log \left(\frac{N-1}{N}i + 1 \right)}{\log(N)} \right\rceil. 
\end{equation}
Changing the base of the logarithm to $N$, one obtains 
\begin{equation} 
l = \left\lceil \log_N \left(\frac{N-1}{N}i + 1 \right) \right\rceil. 
\end{equation}
The same conclusion has been reached \cite{Sudan2006a} but lacking a detailed explanation like ours. 
The case $N = 1$ is trivial, one has $l = i$. Therefore, we conclude that the optimal coding with non-singular codes yields that the length of the $i$-th most probable type is 
\begin{equation}
l_i = \left\{
         \begin{array}{ll}     
         \left\lceil \log_N \left(\frac{N-1}{N}i + 1 \right) \right\rceil & \mbox{~for~} N > 1 \\ 
         i & \mbox{~for~} N = 1. \\
         \end{array}
      \right.
\end{equation}

\section{The relationship between $\Lambda$ and $\tau$}

Here we investigate the consequences of swapping $p_i$ and $p_j$ or $l_i$ and $l_j$ on $\Delta_\Lambda$ and $\Delta_\tau$, the discrete derivative of $\Lambda$ and $\tau$, respectively. For mathematical simplicity, we assume that there are no ties. Then $i$ and $j$ define a concordant pair or a discordant pair. 
Section \ref{derivative_of_concordant_pairs_section} presents a result on the discrete derivative of $n_c$ which is crucial to
conclude in Section \ref{derivative_of_tau_section} that
\begin{itemize}
\item
If the pair $(p_i,l_i)$ and $(p_j, l_j)$ is concordant, $\Delta_\tau < 0$. 
\item
If the pair $(p_i,l_i)$ and $(p_j, l_j)$ is discordant, $\Delta_\tau > 0$.   
\end{itemize}
Moreover, Section \ref{derivative_of_tau_section} also shows that
\begin{itemize}
\item
If the pair $(p_i,l_i)$ and $(p_j, l_j)$ is concordant, $\Delta_\Lambda < 0$. 
\item
If the pair $(p_i,l_i)$ and $(p_j, l_j)$ is discordant, $\Delta_\Lambda > 0$.  
\end{itemize}

\subsection{The discrete derivative of the number of concordant pairs}

\label{derivative_of_concordant_pairs_section}

Hereafter we keep $i$ and $j$ for the subindices of the pairs being swapped and use $x$ and $y$ for the subindices of pairs in general.   
$n_c$, the number of concordant pairs, can be defined as a summation over all pairs, i.e.
\begin{equation}
n_c = \sum_{x=1}^V \sum_{y=1}^V c_{xy},
\end{equation}
where $c_{xy}$ is an indicator variable. $c_{xy} = 1$ if $p_x < p_y$ and $l_x < l_y$ for the $x$-th and the $y$-th type; $c_{xy} = 0$
otherwise. Thus, the pair $(p_x,l_x)$ and $(p_y, l_y)$ is concordant if and only if $c_{xy}$ or $c_{yx}$.
$c_{xy}$ can be expressed as a product of indicator variables, i.e. 
\begin{equation}
c_{xy}= a_{xy} b_{xy},
\end{equation}
where $a_{xy}$ indicates if $p_x < p_y$ and $b_{xy}$ indicates if $l_x < l_y$.

The assumption that there are no ties gives a couple of valuable properties: 
\begin{itemize}
\item
If $x = y$, $a_{xy} = b_{xy} = 0$. 
\item 
If $x \neq y$  
\begin{eqnarray}
a_{xy} = 1 - a_{yx} \\
b_{xy} = 1 - b_{yx} \label{symmetry_equation} 
\end{eqnarray}
by symmetry.
\end{itemize}

We define $n'_c$ as the number of concordant pairs after the swap. $A'$ and $B'$ are the state of matrices $A = \{a_{xy}\}$ and $B = \{b_{xy}\}$ after the swap.
 
\subsection{$l_i$ and $l_j$ are swapped}

If only $l_i$ and $l_j$ are swapped, then $A'= A$, but $B' = B$ is not warranted. Interestingly, the changes in $B$ concern only the $i$-th and the $j$-th row and the $i$-th and the $j$-th column, i.e. 
\begin{equation}
b'_{xy} = \left\{ 
            \begin{array}{ll}
               b_{ji} & \mbox{~if~} x \neq y \mbox{~and~} x = i \mbox{~and~} y = j \\
               b_{ij} & \mbox{~if~} x \neq y \mbox{~and~} x = j \mbox{~and~} y = i \\
               b_{jy} & \mbox{~if~} x \neq y \mbox{~and~} x = i \mbox{~and~} y \neq j \\
               b_{iy} & \mbox{~if~} x \neq y \mbox{~and~} x = j \mbox{~and~} y \neq i \\

               b_{xj} & \mbox{~if~} x \neq y \mbox{~and~} x \neq j \mbox{~and~} y = i \\
               b_{xi} & \mbox{~if~} x \neq y \mbox{~and~} x \neq i \mbox{~and~} y = j \\

               b_{xy} & \mbox{otherwise} \\
            \end{array} 
         \right.
\label{new_b_equation}
\end{equation}  
Then, the 1st derivative of $n_c$ as a function of the number of swaps performed is  
\begin{equation}
\Delta_{n_c} = n'_c - n_c. 
\end{equation}
Suppose that $\gamma$ is sum of the values in the $i$-th and $j$-th row as well as in the $i$-th and $j$-th column of $C = \{c_{ij} \}$ (if a value is found in both a column and a row it will be summed only once) 
and $\gamma'$ as the value of $\gamma$ after the swap. 
Then the 1st derivative can also be defined as
\begin{equation}
\Delta_{n_c} = \gamma' - \gamma.
\end{equation}

By definition,  
\begin{equation}
\gamma = S_1 + S_2 + S_3 + S_4 - T 
\label{raw_loss_equation}
\end{equation}
with 
\begin{eqnarray}
S_1 = \sum_{y=1}^V c_{iy} = \sum_{y=1}^V a_{iy}b_{iy} \\
S_2 = \sum_{y=1}^V c_{jy} = \sum_{y=1}^V a_{jy}b_{jy} \\ 
S_3 = \sum_{x=1}^V c_{xi} = \sum_{x=1}^V a_{xi}b_{xi} \\ 
S_4 = \sum_{x=1}^V c_{xj} = \sum_{x=1}^V a_{xj}b_{xj}
\end{eqnarray}
and
\begin{eqnarray}
T & = & c_{ij} + c_{ji} + c_{ii} + c_{jj}. \nonumber \\
  & = & a_{ij}b_{ij} + a_{ji}b_{ji} + a_{ii}b_{ii} + a_{jj}b_{jj}. 
\end{eqnarray}
$T$ is the sum of the values that have been summed twice by $S_1$, $S_2$, $S_3$ and $S_4$ in Eq. \ref{raw_loss_equation}.
Notice that $S_1+S_2+S_3+S_4 = 2T$ when $V = 2$.
Since $a_{xx}=b_{xx}=0$, 
\begin{equation}
T = a_{ij}b_{ij} + a_{ji}b_{ji}. 
\end{equation}

Recalling Eq. \ref{symmetry_equation}, $S_3$ can be expressed as 
\begin{eqnarray}
S_3 & = & \sum_{y=1}^V (1-a_{iy})(1-b_{iy}) - (1-a_{ii})(1-b_{ii}) \nonumber \\ 
    & = & \sum_{y=1}^V (1-a_{iy})(1-b_{iy}) - 1. 
\end{eqnarray}
Similarly, $S_4$ can be expressed as 
\begin{eqnarray}
S_4 & = & \sum_{y=1}^V (1-a_{jy})(1-b_{jy}) - (1-a_{jj})(1-b_{jj}) \nonumber \\
    & = & \sum_{y=1}^V (1-a_{jy})(1-b_{jy}) - 1.
\end{eqnarray}
On the one hand,
\begin{equation}
S_1 + S_3 = \sum_{y=1}^V a_{iy} (2b_{iy}-1) + V - \sum_{y=1}^V b_{iy} - 1. 
\end{equation}
On the other hand,
\begin{equation}
S_2 + S_4 = \sum_{y=1}^V a_{jy} (2b_{jy}-1) + V - \sum_{y=1}^V b_{jy} - 1. 
\end{equation}
Thus, Eq. \ref{raw_loss_equation} becomes
\begin{eqnarray}
\gamma & = & \sum_{y=1}^V \left[a_{iy} (2b_{iy}-1) + a_{jy} (2b_{jy}-1) \right] - \sum_{y=1}^V (b_{iy} + b_{jy}) + \nonumber \\ 
       &   & 2(V-1)  - a_{ij}b_{ij} - a_{ji}b_{ji}.  \label{loss_equation} 
\end{eqnarray} 

By definition, 
\begin{equation}
\gamma' = S'_1 + S'_2 + S'_3 + S'_4 - T'
\label{raw_gain_equation}
\end{equation}
with
\begin{eqnarray}
S'_1 = \sum_{y=1}^V c'_{iy} = \sum_{y=1}^V a_{iy}b'_{iy} \\ 
S'_2 = \sum_{y=1}^V c'_{jy} = \sum_{y=1}^V a_{jy}b'_{jy} \\ 
S'_3 = \sum_{x=1}^V c'_{xi} = \sum_{x=1}^V a_{xi}b'_{xi} \\ 
S'_4 = \sum_{x=1}^V c'_{xj} = \sum_{x=1}^V a_{xj}b'_{xj}
\end{eqnarray}
and
\begin{eqnarray}
T' & = & c'_{ij} + c'_{ji} + c'_{ii} + c'_{jj} \nonumber \\
   & = & a_{ij}b'_{ij} + a_{ji}b'_{ji} + a_{ii}b'_{ii} + a_{jj}b'_{jj}.
\end{eqnarray}
$T'$ is the sum of the values that have been summed twice by $S'_1$, $S'_2$, $S'_3$ and $S'_4$ in Eq. \ref{raw_gain_equation}.
Notice that $S'_1+S'_2+S'_3+S'_4 = 2T'$ when $V = 2$.
Applying Eq. \ref{new_b_equation}, $T'$ becomes
\begin{eqnarray}
T' & = & a_{ij}b_{ji} + a_{ji}b_{ij} + a_{ii}b_{ii} + a_{jj}b_{jj} \nonumber \\ 
   & = & a_{ij}b_{ji} + a_{ji}b_{ij} \mbox{~~(applying~} a_{xx}=b_{xx}=0 \mbox{)} \nonumber \\
   & = & a_{ij} + b_{ij} - 2a_{ij}b_{ij}. \mbox{~~(applying~} a_{ji} = 1 - a_{ij} \mbox{~and~}b_{ji} = 1 - b_{ij} \mbox{)} \label{redundancy_equation}
\end{eqnarray}

Applying Eq. \ref{new_b_equation}, $S'_1$ can be expressed as 
\begin{eqnarray}
S'_1 & = & \sum_{y=1}^V a_{iy}b_{jy} - a_{ii}b_{ji} - a_{ij}b_{jj} + a_{ii}b_{ii} + a_{ij}(1-b_{ij}) \mbox{~~(applying~} b_{ji} = 1 - b_{ij} \mbox{)} \nonumber \\
     & = & \sum_{y=1}^V a_{iy}b_{jy} + a_{ij}(1-b_{ij})\mbox{~~(applying~} a_{xx}=b_{xx}=0 \mbox{)}
\end{eqnarray}
while $S'_2$ can be expressed as 
\begin{eqnarray}
S'_2 & = & \sum_{y=1}^V a_{jy}b_{iy} - a_{ji}b_{ii} - a_{jj}b_{ij} + a_{ji}b_{ij} + a_{jj}b_{jj} \nonumber \\
     & = & \sum_{y=1}^V a_{jy}b_{iy} + (1 - a_{ij})b_{ij}.
\end{eqnarray}
Similar arguments give
\begin{eqnarray}
S'_3 & = & \sum_{x=1}^V a_{xi}b_{xj} - a_{ii}b_{ij} - a_{ji}b_{jj} + a_{ii}b_{ii} + a_{ji}b_{ij} \nonumber \\ 
     & = & \sum_{x=1}^V a_{xi}b_{xj} + (1 - a_{ij})b_{ij} \\
S'_4 & = & \sum_{x=1}^V a_{xj}b_{xi} - a_{ij}b_{ii} - a_{jj}b_{ji} + a_{ij}b_{ji} + a_{jj}b_{jj} \nonumber \\
     & = & \sum_{x=1}^V a_{xj}b_{xi} + a_{ij} (1 - b_{ij}).
\end{eqnarray}

Recalling Eq. \ref{symmetry_equation}, $S'_3$ can be expressed as 
\begin{eqnarray}
S'_3 & = & \sum_{y=1}^V (1-a_{iy})(1-b_{jy}) - (1-a_{ii})(1-b_{ji}) - (1-a_{ij})(1-b_{jj}) + \nonumber \\ 
    &   & (1 - a_{ij})b_{ij} \\
    & = & \sum_{y=1}^V (1-a_{iy})(1-b_{jy}) - b_{ij} - (1-a_{ij}) + b_{ij} - a_{ij}b_{ij} \nonumber \\
    & = & \sum_{y=1}^V (1-a_{iy})(1-b_{jy}) + a_{ij} - a_{ij}b_{ij} - 1. 
\end{eqnarray}
Similarly, $S'_4$ can be expressed as 
\begin{eqnarray}
S'_4 & = & \sum_{y=1}^V (1-a_{jy})(1-b_{iy}) - (1-a_{ji})(1-b_{ii}) - (1-a_{jj})(1-b_{ij}) + \nonumber \\ 
    &   & a_{ij}(1 - b_{ij}) \nonumber \\ 
    & = & \sum_{y=1}^V (1-a_{jy})(1-b_{iy}) - a_{ij} - (1-b_{ij}) + a_{ij} - a_{ij} b_{ij}  \nonumber \\
    & = & \sum_{y=1}^V (1-a_{jy})(1-b_{iy}) + b_{ij} - a_{ij}b_{ij} - 1. 
\end{eqnarray}

On the one hand,
\begin{equation}
S'_1 + S'_3 = \sum_{y=1}^V a_{iy} (2b_{jy}-1) + V - \sum_{y=1}^V b_{jy} + s'_{1,3} 
\label{first_sum_equation}
\end{equation}
with 
\begin{eqnarray}
s'_{1,3} & = & a_{ij}(1-b_{ij}) + a_{ij} + b_{ji} - 2 \nonumber \\ 
         & = & a_{ij}(1-b_{ij}) + a_{ij} + 1 - b_{ij} - 2 \nonumber \\
         & = & 2a_{ij} - b_{ij} - a_{ij}b_{ij} - 1.
\end{eqnarray}
On the other hand,
\begin{equation}
S'_2 + S'_4 = \sum_{y=1}^V a_{jy} (2b_{iy}-1) + V - \sum_{y=1}^V b_{iy} + s'_{2,4} 
\label{second_sum_equation}
\end{equation}
with 
\begin{eqnarray}
s'_{2,4} & = & a_{ji}(1-b_{ji}) + a_{ji} + b_{ij} - 2 \nonumber \\
         & = & (1-a_{ij})b_{ij} + 1 - a_{ij} + b_{ij} - 2 \nonumber \\
         & = & - a_{ij} + 2b_{ij} - a_{ij}b_{ij} - 1.  
\end{eqnarray}
Finally, Eqs. \ref{redundancy_equation}, \ref{first_sum_equation} and \ref{second_sum_equation} transform Eq. \ref{raw_gain_equation} into
\begin{eqnarray}
\gamma' & = & \sum_{y=1}^V \left[a_{iy} (2b_{jy}-1) + a_{jy} (2b_{iy}-1) \right] - \sum_{y=1}^V (b_{iy} + b_{jy}) + \nonumber \\ 
      &  & 2(V - a_{ij}b_{ij} - 1) + a_{ij} + b_{ij}
\label{gain_equation}
\end{eqnarray}

Combining Eqs. \ref{loss_equation} and \ref{gain_equation} yields, after some algebra,
\begin{eqnarray}
\Delta_{n_c} & = & \gamma' - \gamma \nonumber \\
       & = & 2\sum_{y=1}^V \alpha_y \beta_y + 1 \label{discrete_derivative_equation}
\end{eqnarray}  
where 
\begin{eqnarray}
\alpha_y & = & a_{iy} - a_{jy} \\ 
\beta_y  & = & b_{jy} - b_{iy}. 
\end{eqnarray}

The final formulae for $\gamma$ (Eq. \ref{loss_equation}), $\gamma'$ (Eq. \ref{gain_equation}) and $\Delta_{n_c}$ (Eq. \ref{discrete_derivative_equation}) have been verified with the help of computer simulations. For a given $V$, the simulation is based on the following algorithm: 
\begin{itemize} 
\item
Setup
  \begin{enumerate}
  \item
  Generate a vector $A$ of size $V$ containing numbers from $1$ to $V$ (in that order).  
  \item
  Generate a vector $B$ of size $V$ containing numbers from $1$ to $V$ (in that order) if the initial state is $\tau = 1$ and containing those numbers in the reverse order if that state is $\tau = -1$. 
  \item
  Calculate $n_c$ and $\gamma$ with $A$ and $B$.
  \end{enumerate}  
\item
Test: run $T$ times  
  \begin{enumerate}
  \item 
  Choose uniformly at random two integers $i$ and $j$ such that $1 \leq i, j \leq V$ and $i\neq j$.
  \item 
  Swap the $i$-th and the $j$-th element of $B$.
  \item  
  Calculate $n_c'$ and $\gamma'$ with $A$ and the new $B$. 
  \item
  Check that 
     \begin{enumerate}
     \item
     $\gamma$ coincides with the value provided by Eq. \ref{loss_equation}.
     \item
     $\gamma'$ coincides with the value provided by Eq. \ref{gain_equation}.
     \item
     $\Delta_{n_c} = n_c'-n_c$ coincides with the value provided by Eq. \ref{discrete_derivative_equation}.
     \end{enumerate}
  \item
  $n_c = n_c'$, $\gamma = \gamma'$.
  \end{enumerate} 
\end{itemize}
The algorithm was run successfully for $V = 1$ to $100$ with $T = 10^5$ and both initial states. 


\subsubsection{$p_i$ and $p_j$ are swapped}

Notice that this case is equivalent to the case when $l_i$ and $l_j$ are swapped by symmetry. It suffices to exchange the role of the matrices $A$ and $B$. 
Now $a_{xy}$ indicates if $l_x < l_y$ and $b_{xy}$ indicates if $p_x < p_y$. 

\subsubsection{The variation of $\tau$}

\label{derivative_of_tau_section}

If ties are missing, $n_d = n_0 - n_c$ and $\tau$ becomes  
\begin{equation}
\tau = \frac{2 n_c}{n_0} - 1. 
\end{equation}
Then the discrete derivative of $\tau$ (as a function of the number of swaps) is 
\begin{eqnarray}
\Delta_{\tau} & = & \frac{n_c' - n_c}{n_0} \\
              & = & \frac{\Delta_{n_c}}{n_0},
\end{eqnarray} 
where $\Delta_{n_c}$, the discrete derivative of $n_c$, satisfies   
\begin{equation}
\Delta_{n_c} = 2\sum_{y=1}^V \alpha_y \beta_y + 1
\end{equation}  
as explained in Section \ref{derivative_of_concordant_pairs_section}.

Without any loss of generality, suppose that $i < j$. We want to prove that 
\begin{itemize}
\item[{\em Statement 1}]
If the pair $(p_i,l_i)$ and $(p_j, l_j)$ is concordant then $\Delta_{n_c} < 0$, which is equivalent to 
\begin{equation}
\sum_{y=1}^V \alpha_y \beta_y < -\frac{1}{2}.
\end{equation}
\item[{\em Statement 2}]
If $i$ and $j$ are a discordant pair then $\Delta_{n_c} > 0$, which is equivalent to 
\begin{equation}
\sum_{y=1}^V \alpha_y \beta_y > -\frac{1}{2}.
\end{equation}
\end{itemize}
First, we notice some relevant properties of $\alpha$ and $\beta$. Suppose that $p_1 < p_2 < ... < p_k < ... < p_V$. Then it is easy to see that 
\begin{itemize}
\item[{\em Property 1}]
$\alpha_y = 1$ if $i < y < j$ and $\alpha_y = 0$ otherwise. 
\end{itemize}
Now suppose that $l_1 < l_2 < ... < l_k < ... < l_V$. Then it is easy to see that
\begin{itemize}
\item[{\em Property 2}]
$\beta_y = -1$ if $i < y < j$ and $\beta_y = 0$ otherwise. 
\end{itemize}
If the pair $(p_i,l_i)$ and $(p_j, l_j)$ $i$ is concordant, Properties 1 and 2 indicate that $\alpha_y \beta_y \in \{-1, 0\}$, giving 
\begin{equation}
\sum_{y=1}^V \alpha_y \beta_y \leq 0.
\end{equation}
Adding that $\alpha_i \beta_i = - a_{ji}b_{ji} = -1$ or $\alpha_j \beta_j = - a_{ij}b_{ij} = -1$ when the pair $(p_i,l_i)$ and $(p_j, l_j)$ is concordant, it follows that
\begin{equation}
\sum_{y=1}^V \alpha_y \beta_y \leq -1,
\end{equation}
which proves Statement 1. 
If $i$ and $j$ are discordant, Properties 1 and 2 indicate that $\alpha_y \beta_y \in \{0,1\}$, giving 
\begin{equation}
\sum_{y=1}^V \alpha_y \beta_y \geq 0,
\end{equation}
which proves Statement 2.

\subsubsection{The variation of $\Lambda$}

\label{derivative_of_cost_section}

The value of $\Lambda$ after one of those swaps will be examined considering two cases. 
First, imagine that $l_i$ and $l_j$ are swapped. The value of $\Lambda$ after the swap is
\begin{equation}
\Lambda' = \Lambda - (p_i \lambda_i + p_j \lambda_j) + (p_i \lambda_i' + p_j \lambda_j').
\end{equation}
Applying $\lambda_i' = \lambda_j$ and $\lambda_j' = \lambda_i$ it is obtained
\begin{equation}
\Delta_\Lambda = (p_i - p_j) (g(l_j) - g(l_i)). \label{swap_of_lengths_equation}
\end{equation}  
The fact that $g(l)$ is a monotonically increasing function of $l$ allows one to conclude that
\begin{itemize}
\item
If the pair was concordant then $\Delta_\Lambda < 0$. 
\item
If the pair was discordant then $\Delta_\Lambda > 0$. 
\end{itemize}

Second, imagine that $p_i$ and $p_j$ are swapped. The value of $\Lambda$ after the swap is
\begin{equation}
\Lambda' = \Lambda - (p_i \lambda_i + p_j \lambda_j) + (p_i' \lambda_i + p_j' \lambda_j).
\end{equation}
Applying $p_i' = p_j$ and $p_j' = p_i$ it is obtained again Eq. \ref{swap_of_lengths_equation}. Thus, the same arguments and conclusions apply to the swap of $p_i$ and $p_j$.

\section{The constancy conditions}

\subsection{Preliminaries: a lower bound for $E[\lambda]$}

\label{lower_bound_subsection}

The optimal coding method presented above allows one to derive a lower bound for 
\begin{equation}
E[\lambda] = \frac{1}{V} \sum_{i=1}^{V} \lambda_i. 
\label{definition_equation}
\end{equation}
The point is that $E[\lambda]$ can be seen as a particular case of $\Lambda$ with $p_i = 1/V$. Suppose that $l_{max}$ is the maximum string length needed by the optimal coding above.
Obviously, $l_{max}$ is the smallest integer such that 
\begin{equation}
V \leq \sum_{l=1}^{l_{max}} N^l.
\end{equation}
Such an optimal coding requires all strings of length smaller than $l_{max}$ and
\begin{equation}
U = V - \sum_{l=1}^{l_{max - 1}} N^l 
\end{equation}
strings of length $l_{max}$. 
Therefore, Eq. \ref{definition_equation} gives 
\begin{equation}
E[\lambda] \geq \frac{1}{V} \left( \sum_{l=1}^{l_{max}-1} g(l) \right) + U g(l_{max}). 
\label{lower_bound_equation}
\end{equation} 

\subsection{The constancy of $V$, $E[\lambda]$, $\sigma[\lambda]$ and $\sigma[p]$}

The constancy of $V$ could derive from the existence of a core lexicon \cite{Cocho2015a,Gerlach2013a,Petersen2012b,Ferrer2000a}
and social constraints on the addition and propagation of new types \cite{Baronchelli2005a}. 

The constancy of $E[\lambda]$ and $\sigma[\lambda]$ could be due to cost-cutting pressures. $V E[\lambda]$ can be regarded as the cost of learning the strings of symbols making the repertoire and storing them in memory. Suppose that every type is assigned a different string of symbols, i.e. the coding scheme is nonsingular \cite{Cover2006a}. One could use strings of length $\lceil \log_N V \rceil$ to code for every type but this would be a waste. A lower bound for $E[\lambda]$ is given by Eq. \ref{lower_bound_equation}.
We expect that natural systems are attracted towards this lower bound to minimise the cost of storing the repertoire, providing support for the simplifying assumption that $E[\lambda]$ is constant.
  
The constancy of $\sigma[\lambda]$ can be supported by the need of intermediate values of $\sigma[\lambda]$: 
\begin{itemize}
\item
A small value of $\sigma[\lambda]$ might be difficult or impossible to achieve. Let us consider that $\sigma[\lambda]$ is minimum, i.e. $\sigma[\lambda] = 0$, which is equivalent to all types having the same magnitude $k$. If $V < N^k$ then some types are not distinguishable (the coding scheme is nonsingular), which is something to avoid.
Recall that the unconstrained solution to the minimisation of $\Lambda$, i.e. Eq. \ref{trivial_minimization_equation} yields $\sigma[\lambda] = 0$ but sacrificing the distinguishability of types (if $V>N$, distinguishability imposes that $\sigma[\lambda]$ is bounded below by a non-zero value). Although it is possible to code any repertoire with strings of length 1 from an alphabet (one only needs that $V=N$), strings of length greater than one have been shown to be evolutionary advantageous to combat noise \cite{Plotkin2000}.   
If $V \geq N^k$ then the high cost implied by the value of $E[\lambda]$ (as explained above) turns $\sigma[\lambda] = 0$ unlikely. 
A further reason against $\sigma[\lambda] = 0$ is that, under pressure to minimise $E[\lambda]$ or $\Lambda$ close to the optimal coding for nonsingular codes, all lengths up to $l_{max}$ are taken. 
\item
Let us consider that $\sigma[\lambda]$ is large (much greater than the value of $\sigma[\lambda]$ for nonsingular codes). Then very long strings are expected but this is unnecessarily costly and then less likely to happen. 
\end{itemize}
Finally, the constancy of $\sigma[p]$ could arise from mechanisms that shape symbol probabilities independently from $\Lambda$ but that can still involve cost-cutting factors \cite{Ferrer2004e,Newman2004a}.

\section{Optimisation and correlation in community detection}

We have shown above an intimate relationship between $\Lambda$ and $r$. 
A straightforward relationship between a function to optimise and correlation is also found in methods for community detection in networks \cite{Fortunato2010a}. A central concept in those methods is $Q$, a measure of the quality of a partitioning into communities of a network that must be maximised \cite{Fortunato2010a}. 
Interestingly, $Q$ is intimately related with Fisher's intraclass correlation \cite{Koch1982a}, another correlation coefficient that should not be confused with the popular Pearson interclass  correlation $r$ that we have considered above. To see it in detail, suppose that $m$ is the number of edges of a network, $k_i$ is the degree of the i-th vertex, $A$ is the adjacency matrix and $c_i$ is the community to which the $i$-th vertex belongs, $Q$ can be defined as \cite[p. 224]{Newman2010a}
\begin{equation}
Q = \frac{1}{2m} \sum_{i}\sum_{j} \left(A_{ij} - \frac{k_i k_j}{2m} \right) \delta(c_i, c_j),
\end{equation}
where $\delta(x,y)$ is the Kronecker delta ($\delta(x,y) = 1$ if $x = y$; $\delta(x,y) = 0$ otherwise).
The intraclass correlation that is connected with $Q$ is defined between the communities at both ends of an edge. If $x_i$ is a scalar quantity associated to the $i$-th vertex, the intraclass correlation between $x_i$ and $x_j$ over edges is $r_{intra} = cov(x_i, x_j)/\sigma_x^2$,
where $\sigma_x$ is the standard deviation of the $x_i$'s and $cov_{intra}(x_i, x_j)$ is the intraclass covariance between $x_i$ and $x_j$ over edges, i.e. \cite[p. 228]{Newman2010a}     
\begin{equation}
cov_{intra}(x_i, x_j) = \frac{1}{2m} \sum_i \sum_j \left(A_{ij} - \frac{k_i k_j}{2m}\right) x_i x_j. 
\end{equation} 
The similarity between $Q$ and $cov_{intra}(x_i, x_j)$ is strong: the only difference is that $\delta(c_i, c_j)$ in $Q$ is replaced by $x_i x_j$ in $cov_{intra}(x_i, x_j)$ \cite[p. 228]{Newman2010a}. For the case of only two communities, one may define a variant of $Q$, namely 
\begin{equation}
Q' = \frac{1}{2m} \sum_{i}\sum_{j} \left(A_{ij} - \frac{k_i k_j}{2m} \right) s_i s_j,
\end{equation}
where $s_i \in \{-1, +1 \}$, indicates the community of the $i$-th vertex. Then, $Q' = cov(s_i, s_j)$. Thus, the maximisation of $Q'$ is fully equivalent to the maximisation of $r_{intra}$.

\section{Applications beyond the law of abbreviation}
   
The results presented in this article go beyond Zipf's law of abbreviation. For instance, the online memory cost of a sentence of $n$ words can be
defined as \cite{Ferrer2013e,Ferrer2014c}
\begin{equation}
D = (n - 1) \sum_{d=1}^{n-1}p(d)g(d),
\end{equation}
where $n - 1$ is the number of edges of a syntactic dependency tree of $n$ vertices,
$p(d)$ is the proportion of dependencies of length $d$ and $g(d)$ is the cognitive cost
of a dependency of length $d$. 
Recently, it has been argued that $g(d)$ may not be a monotonically decreasing function of $d$ as commonly believed \cite{Alday2015a}.
The minimisation of $D$ can be regarded as particular case of $\Lambda$ where $V = n - 1$, $p_i$ is $p(d)$ and $g(l_i)$ is $g(d)$ and thus a negative correlation between $p(d)$ and $g(d)$ is predicted applying the arguments employed in this article. Finally, knowing that $p(d)$ is a decreasing function of $d$ in real syntactic dependencies \cite{Ferrer2004b,Liu2007a} and under the null hypothesis that the words of a sentence are arranged linearly at random \cite{Ferrer2004b}, a positive correlation between $d$ and $g(d)$ follows.

\end{document}